\documentclass[twocolumn,showpacs,amsmath,amssymb,prb,graphics,graphicx
]{revtex4}
\usepackage{bm}
\usepackage{graphicx}
\usepackage{dcolumn}
\begin{document}
\title
{Meissner response of anisotropic superconductors}
\author{V. G. Kogan}
\affiliation{Ames Laboratory DOE and Physics Department ISU, Ames, IA 50011}

\begin{abstract}
 The response field of a half-space anisotropic superconductor is
evaluated for an arbitrary weak external field source. Example sources  of a
 point magnetic moment and a circular current are considered in
detail. For the penetration depth $\lambda \ll L$ with $L$ being any other
relevant distance (the source size, or the distance between the source and the
superconductor), the major contribution to the response  is 
the $\lambda$ independent field of the source image. It is shown that the
absolute value of $\lambda$ cannot be extracted from the response field with a
better accuracy than that for the source position. Similar problems are
considered for thin films. 
  \end{abstract}
\maketitle

\section{Introduction}

 A new experimental technique, the scanning SQUID microscopy (SSM), has
recently been developed for measuring magnetic field distributions due to 
 vortices exiting superconducting samples.\cite{Kirtley} Knowing the
distributions  one can, in principle, extract the London penetration depth
$\lambda$ (either isotropic or anisotropic) and its temperature
dependence.\cite{KirtleyT} In other implementation of this method, one can 
  measure the Meissner response of a superconductor to a
weak external source of static magnetic field. Again, the response  may provide
information about the $\lambda$ temperature dependence and its anisotropy. Also,
one can use tunable external field sources to study elementary forces acting upon
vortices exiting the surface.\cite{Moler-Gardner} Similar problems are
encountered in the magnetic force microscopy applied to surfaces of anisotropic
superconductors. 

There are a few publications addressing these problems for 
  isotropic superconductors.\cite{Clem,Milosevic,Obukhov} For
anisotropic materials, however, the response field is asymmetric even when the 
source has certain symmetries, and one cannot use methods developed for  
isotropic materials. To deal with the problem, one can utilize the
two-dimensional (2D) Fourier transform with respect to coordinates $x,y$ of the
interface, provided the  equations for the field distributions inside and outside
the superconducting half-space are {\it linear}. This is the case if one adopts
the London description for the field inside the material. Then, one solves the
remaining system of ordinary differential equations in the variable 
$z$, normal to the interface. This approach has been developed in Ref.
\onlinecite{KSL} for the problem of vortices crossing the superconductor surface. 

Let us consider a source with known field distribution   ${\bm
h}^s $ in the absence of superconductor. 
In the presence of the superconductor occupying the  half-space
$z<0$, the total field in vacuum $z>0$ can be written as 
\begin{equation}
{\bm h} = {\bm h}^s + {\bm h}^r\,,
\end{equation}
 where ${\bm h}^r$ is the response field which satisfies div$\,{\bm
h}^r=$ curl$\,{\bm h}^r=0$ in vacuum. One can look for this field as
$\nabla\varphi^r$ with the potential $\varphi^r$ obeying the Laplace equation
and the zero boundary condition far from the surface. The general form of such
a potential is
\begin{equation}
\varphi^r ({\bm r},z)=\int\frac{d^2{\bm k}}{(2\pi)^2}\,\varphi^r ({\bm k})\,
e^{i{\bm k}\cdot{\bm r}-kz}\,.
\label{Laplace}
\end{equation}
  Here, ${\bm r}=(x,y)$ and $z$ is directed normal to the superconducting flat
surface at $z=0$; $\varphi^r ({\bm k})e^{-kz}$ is the  2D Fourier
transform with respect to variables $x,y$ at any fixed  $z>0$. The potential
(\ref{Laplace}) is defined only in the upper half-space; hence, there is no
problem of uniqueness which is in general associated with the description of the
magnetic field by a potential. 

The field inside the superconductor satisfies London equations which read
in general anisotropic case as\cite{K81} 
\begin{equation}
h_i-\lambda^2m_{lk}\,e_{lst}\,e_{kni}\,  h_{t,ns}= 0 \,,\quad i=x,y,z\,.
\label{general}
\end{equation}
Here $e_{ikl}$ is the unit antisymmetric tensor, $m_{ij}$  is the mass
tensor, $h_{t,ns}$ abbreviates $\partial^2h_t/\partial x_n\partial x_s$. 
The average penetration depth $\lambda =(\lambda_a\lambda_b\lambda_c)^{1/3}$ is
related to the actual penetration depth for the currents, e.g., along the
crystal direction $a$: $\lambda_a=\lambda\sqrt{m_a}$. The masses are normalized
so that $m_am_bm_c=1$. 

The problem, therefore, is to match the solutions for the field inside and outside
the superconductor with boundary conditions of the field continuity at the
interface. 

\section{Response field}

Usually in situations of interest, the sample surface is normal to one of the
principal crystal directions. We call this direction $c$ and choose  the frame   
$x,y,z$ as coinciding with $a,b,c$.   In this situation, the mass
tensor is diagonal ($m_{xx}= m_a,\quad m_{yy}=m_b,\quad  m_{zz}= m_c$),
and Eqs. (\ref{general}) reduce to
\begin{eqnarray}
h_x
&+&\lambda_b^2(h_{z,x}-h_{x,z})_{,z}+\lambda_c^2(h_{y,x}-h_{x,y})_{,y}=0\,
,\nonumber\\
h_y &+&\lambda_c^2(h_{x,y}-h_{y,x})_{,x}+\lambda_a^2(h_{z,y}-h_{y,z})_{,z}=0\,,
\label{London_system}\\
h_z
&+&\lambda_a^2(h_{y,z}-h_{z,y})_{,y}+\lambda_b^2\,(h_{x,z}-h_{z,x})_{,x}=0\,.
\nonumber
\end{eqnarray}
One can replace any of one of these equations by div${\bm
h}=h_{x,x}+h_{y,y}+h_{z,z}=0$. It is convenient to replace the third one, and
exclude $h_{z,z}$ from the first two:
\begin{eqnarray}
\noindent  h_x-\lambda_b^2h_{x,xx}&-&\lambda_c^2h_{x,yy} -
\lambda_b^2h_{x,zz}+(\lambda_c^2-\lambda_b^2)h_{y,xy} =0,\nonumber\\ 
\noindent  h_y-\lambda_c^2h_{y,xx}&-&\lambda_a^2h_{y,yy} -
\lambda_a^2h_{y,zz}+(\lambda_c^2-\lambda_a^2)h_{x,xy} =0,\nonumber\\
   h_{z,z}= &-&\,(h_{x,x}+h_{y,y}) \,.
\label{London_1} 
\end{eqnarray}
Note that the first two equations are decoupled from the third. 
 
We now apply  the $x,y$ Fourier transform to Eqs.
(\ref{London_1}):
\begin{eqnarray}
&&(1+\lambda_b^2k_x^2 +\lambda_c^2k_y^2)h_x -\lambda_b^2 h''_x
+(\lambda_b^2-\lambda_c^2)k_xk_yh_y  =0\,,\nonumber\\
&&(\lambda_a^2-\lambda_c^2)k_xk_yh_x +(1+\lambda_c^2k_x^2 +\lambda_a^2k_y^2)h_y
-\lambda_a^2 h''_y   =0\,,\nonumber\\
&&i(k_xh_x+k_yh_y)+h_z'=0\,. \label{FT}
 \end{eqnarray}
Here,  $ h_i $ are functions of $ k_x,k_y$ and $z$,
and the prime denotes the derivative with respect to $z$. Hence, we are left with
the linear system of ordinary 2nd order differential equations with respect to
the variable $z$ for $h_i({\bm k},z)$.   

The solutions are linear combinations of simple exponentials:
\begin{equation}
h_i({\bm k},z) = \sum_n H_i^{(n)}({\bm k})e^{q_nz} \,.
\label{exp}
\end{equation}
The parameters $q_n$ and their number are to be determined. Substituting each
term of Eq. (\ref{exp}) in the system (\ref{FT}) we obtain a linear
homogeneous system for $H_i^{(n)}$: 
\begin{eqnarray}
\noindent
&&H_x\,(1+\lambda_b^2k_x^2+\lambda_c^2k_y^2-\lambda_b^2q^2)+H_y\,(\lambda_b^2-\lambda_c^2)k_xk_y
 =0,\nonumber\\
\noindent &&H_x\,(\lambda_a^2-\lambda_c^2)k_xk_y
+H_y\,(1+\lambda_c^2k_x^2+\lambda_a^2k_y^2-\lambda_a^2q^2)=0, \nonumber\\
\noindent &&i\,(k_xH_x +k_yH_y)+qH_z=0\,. \label{FT1} 
\end{eqnarray}
for each $n$; the superscript $n$ is omitted for brevity.
As has been mentioned, the first two equations here are decoupled from the third;
they have a nonzero solution provided  their determinant is zero. This gives a
quadratic equation for $q^2$ which is readily solved: 
\begin{eqnarray}
q^2_{1,2}&=&\frac{ P\pm\sqrt{Q}}{2\lambda^2_a\lambda^2_b}
\,, \label{q's}\\ 
P&=&\lambda^2_a+\lambda^2_b+\lambda^2_c(\lambda^2_bk_x^2+\lambda^2_ak_y^2)+
\lambda^2_a\lambda^2_bk^2\,,\nonumber \\
Q&=& P^2
-4\lambda^2_a\lambda^2_b(1+\lambda^2_bk_x^2+\lambda^2_ak_y^2)(1+\lambda^2_ck^2)\,.
\nonumber
\end{eqnarray}
  Note that $Q<P^2$ and both $q^2_1$ and $q^2_2$ are positive;
therefore,  there are only two  positive $q$'s (i.e., $n=1,2$) which satisfy the 
requirement of vanishing fields at $z=-\infty$. 

The quantities $q$ determine how the field attenuates in the superconductor. We
now have to find the ``amplitudes" $H_i$ from Eqs. (\ref{FT1})
for each $q$. It is worth noting that solving the homogeneous system of linear
equations implies, in fact,  expressing some unknowns in terms of others. To
this end, one has to determine the rank of the matrix of coefficients for the
system (\ref{FT1}), choose a proper subsystem to solve, etc.  The actual
procedure might differ depending on the situation in question. 

Solving the system (\ref{FT1}) we obtain:
 \begin{eqnarray}
H_x&=&-i\frac{d-\lambda_a^2q^2}{k_xq\,(\lambda^2_a-\lambda^2_b)}\, H_z
,\quad H_y=i \,\frac{ d-\lambda^2_bq^2}{k_yq\,(\lambda^2_a-\lambda^2_b)}\,H_z
\,,\nonumber\\
d&=&1+\lambda_b^2k_x^2+\lambda_a^2k_y^2\,.\label{Hxy}
\end{eqnarray}
 for each $q$ of Eq. (\ref{q's}). 

Let us turn now to the field ${\bm h}^s$ of the source. As a consequence of
div$\,{\bm h}^s=$ curl$\,{\bm h}^s=0$ out of the source, the 2D Fourier
components of ${\bm h}^s$ are not independent.  As with the response field, we
can look for this field in the from ${\bm h}^s=\nabla\,\varphi^s$ such that
$\nabla^2\varphi^s=0$. In our situation,  the source is situated in the upper
half-space $z>0$, and we are interested in the field ${\bm h}^s$ ``under" the
source. The general solution of the Laplace equation which vanishes as
$z\rightarrow -\infty$ is:
\begin{equation}
\varphi^s ({\bm r},z)=\int\frac{d^2{\bm k}}{(2\pi)^2}\,\varphi^s ({\bm k})\,
e^{i{\bm k}\cdot{\bm r}+kz}\,.
\label{Laplace_s}
\end{equation}
The 2D Fourier components of the source field at the interface then read:
\begin{equation}
 h_{\alpha}^s = i\,k_{\alpha} \varphi^s ({\bm k})\quad (\alpha=x,y)\,,\quad   
h_z^s =k\varphi^s ({\bm k})\,.\label{2dh}
\end{equation}
Hence, the boundary conditions take the form:
\begin{eqnarray}
  i\,k_{\alpha}(\varphi^s+\varphi^r) &=& H_{\alpha}^{(1)} +
H_{\alpha}^{(2)} 
\quad (\alpha=x,y),\nonumber\\   
 k(\varphi^s - \varphi^r) &=& H_z^{(1)} +H_z^{(2)}\,.\label{bc1} 
\end{eqnarray} 
Since the components $H_{\alpha}$ are expressed in terms of $H_z$'s, we can solve 
the system  (\ref{bc1}) to obtain:
  \begin{eqnarray}
\varphi^r &=& \varphi^s \,\frac{ k(q_1+q_2-k)- 
q_1q_2(1-1/d)}{ k(q_1+q_2+k)+ q_1q_2(1-1/d)}\,,\nonumber \\  
H_z^{(1)}&=&\frac{k[\varphi^s(q_2-k)-\varphi^r(q_2+k)]}{q_2-q_1}\,,\label{phi_an}
\\
H_z^{(2)}&=&\frac{k[\varphi^s(k-q_1)+\varphi^r(q_1+k)]}{q_2-q_1}\,.\nonumber
\end{eqnarray}
Thus, the response outside and inside the superconductor is
expressed in terms of the source field $\varphi^s$. It is worth noting that since $\varphi^s({\bm k})$ can be replaced with
$ h_z^s({\bm k})/k$, the response field can be expressed in terms of the
$z$ component of the source field at the interface.

 One can see, in particular, that the total flux in
$z$ direction "reflected" by the superconductor is equal and opposite in
sign to the incident flux of the source crossing the interface. Indeed,
this flux is 
\begin{equation}
h_z^r|_{k=0}=-k\varphi^r|_{k=0} = -k \varphi^s|_{k=0}=-h_z^s|_{k=0} \,.
\end{equation}

It is seen from the general formulas (\ref{Hxy}) that the case
$\lambda_a=\lambda_b$ is singular. The formal reason for this is that the 
first two equations of the system (\ref{FT1}) (which we have used to express
$H_x$ and $H_y$ in terms of $H_z$) are no longer  independent.  This
situation should be treated separately.

\subsection{Small $\bm\lambda$ }

In many situations, the penetration depths are small relative to other relevant
lengths in the problem such as the distance $z_0$ between the source and the
interface. The characteristic $k$'s then satisfy $k\lambda \ll 1$. The relations
between the response and the source fields then simplify. In this
approximation, Eqs. (\ref{q's}) give $q_1=1/\lambda_a$ and $q_2=1/\lambda_b$ for
$\lambda_a<\lambda_b$. Then, Eqs. (\ref{Hxy}) yield $d=1$ and
 \begin{eqnarray}
H_x^{(1)}=0\,,\quad H_z^{(1)}=-ik_y\lambda_aH_y^{(1)}\,,\label{Hy1_sm}\\
H_y^{(2)}=0\,,\quad H_z^{(2)}=-ik_x\lambda_bH_x^{(2)}\,.\label{Hx2_sm}
\end{eqnarray}
Finally, the boundary conditions give:
  \begin{eqnarray}
\varphi^r &=& \varphi^s \,\Big(1-2\,\frac{\lambda_bk_x^2 +\lambda_ak_y^2}{k
}\Big),\label{f_sm}\\  
H_y^{(1)}&=&{k_y\over k_x}H_x^{(2)}=2ik_y\varphi^s \,\Big(1-\frac{\lambda_bk_x^2
+\lambda_ak_y^2}{k }\Big).\label{hx1_sm}    
\label{small_l}
\end{eqnarray}
Note that in this approximation $\lambda_c$ does not enter the result; in other
words, the currents along $z$ are small and can be disregarded. The results 
(\ref{Hy1_sm})-(\ref{hx1_sm}) can also be obtained   directly 
  starting with the London Eqs. (\ref{London_system}) and taking advantage of 
  $\partial/\partial z\gg \partial/\partial x_{\alpha}$. 

It is also worth observing that in zero order in   $k\lambda\ll 1$,
$\varphi^r=\varphi^s$. This means that the $x$ and $y$ components of the response
field are the same as those for the source, whereas $h_z^r=-h_z^s$, see Eqs.
(\ref{Laplace}) and (\ref{Laplace_s}). In other words, in this approximation the
response field is the mirror image of the source field. 

\subsection{Isotropic materials}

It is readily seen that in this case 
\begin{equation}
q_1=q_2=\sqrt{\lambda^{-2}+k^2}\,.\label{q_is}
\end{equation}
The first two of Eqs. (\ref{FT1}) turn identities, whereas the third gives one of
$H_i$'s in terms of two others, e.g., $H_z=(k_xH_x +k_yH_y)/iq$. The boundary
conditions then give:
\begin{equation}
\varphi^r =\frac{q-k }{q+k}\,\varphi^s \,,\quad H_{\alpha}=\frac{2iq
}{q+k}\,k_{\alpha}\,\varphi^s\,. 
\label{isotropic}
\end{equation}

\subsection{${\bm \lambda}_a={\bm\lambda}_b< {\bm\lambda}_c$}

This is the case, e.g., of BSCCO crystal with the $ab$ plane being the surface. 
Equations (\ref{q's}) yield:
\begin{equation}
q_1=\sqrt{\lambda_{ab}^{-2}+k^2}\,,\qquad
q_2=\sqrt{\lambda_{ab}^{-2}+\gamma^2k^2}\,,\label{BSCCO1}
\end{equation}
where $\gamma=\lambda_c/\lambda_{ab}$ is the anisotropy parameter. Substituting
$q_1$ in the first two equations of the system (\ref{FT1}), we obtain two
identical results $H_x^{(1)}k_y-H_y^{(1)}k_x=0$ which together with the third
equation yield:
\begin{equation}
H_x^{(1)}=\frac{iq_1k_x}{k^2}\,H_z^{(1)}\,,\quad
H_y^{(1)}=\frac{iq_1k_y}{k^2}\,H_z^{(1)}\,.\label{BSCCO2}
\end{equation}
Doing the same for $q_2$, we obtain from the first two equations
$H_x^{(2)}k_x+H_y^{(2)}k_y=0$ which is compatible with the third only if
$H_z^{(2)}=0$. Hence, we have:
\begin{equation}
 H_z^{(2)}=0\,,\qquad
H_y^{(2)}=-\,\frac{ k_x}{k_y}\,H_x^{(2)}\,.\label{BSCCO3}
\end{equation}
The boundary conditions (\ref{bc1}) 
yield:
\begin{equation}
\varphi^r =\frac{q_1-k}{q_1+k}\,\varphi^s \,,\quad H_z^{(1)}=\frac{2k^2
}{q_1+k}\, \varphi^s\,,\quad {\bm H}^{(2)}=0\,.\label{BSCCO5}
\end{equation}
Note that for this  case, $q_2$ along with $\lambda_c$
drop off the result for {\it any} source. In other words, the response is as if
the superconductor were isotropic with the penetration depth $\lambda_{ab}$. 

\subsection{${\bm \lambda}_c={\bm\lambda}_a< {\bm\lambda}_b$}

This is the case of  the screening by the ``side surface" of a
uniaxial crystal like BSCCO. Our notation is, however, differs from the commonly
used (for standard notation, we should have replaced in our formulas   
$\lambda_c$ and $\lambda_a$ with 
$\lambda_{ab}$ and $\lambda_b\rightarrow\lambda_c$). We then obtain using Eq.
(\ref{q's}):
\begin{equation}
q_1=\sqrt{\lambda_{a}^{-2}+ k^2}\,,\quad 
q_2=\sqrt{1+\lambda_{b}^2k_x^2+\lambda_a^2k_y^2}\Big/\lambda_b\,.
\label{q_1_2}
\end{equation}
With $q=q_1$, Eq. (\ref{Hxy}) gives
\begin{equation}
H_x^{(1)}=i\,\frac {k_x}{q_1}\,H_z^{(1)}\,,\quad
H_y^{(1)}=i\,\frac{1+\lambda_{a}^2k^2_y}{\lambda_a^2k_yq_1}\,H_z^{(1)}\,,\label{H1}
\end{equation}
and for $q=q_2$
\begin{equation}
 H_x^{(2)}=i\,\frac{q_2}{k_x}\,H_z^{(2)}\,,\qquad
H_y^{(2)}=0\,.\label{H1}
\end{equation}
Finally, Eqs. (\ref{phi_an}) give $\varphi^r$,  $H_z^{(1)}$, and $H_z^{(2)}$ in
terms of the source field $\varphi^s$.  

\section{Examples of sources}

To apply formulas derived above for the response fields outside and inside
superconductor, one needs the Fourier transform $\varphi^s({\bm k})$ of the source
field at the interface. Below we provide  examples for which $\varphi^s({\bm
k})$ can be calculated analytically.

\subsection{Point magnetic moment}

Consider a magnetic moment ${\bm\mu}$ situated at the hight $z_0$ above the
superconductor. The corresponding potential (in the absence of the
superconductor) at $z=0$ is
\begin{equation}
\varphi^s=-\,\frac{{\bm\mu}\cdot   {\bm R}}{R^3} =  \frac{\mu_z\,z_0 -
{{\bm\mu}}\cdot{\bm r}}
 {(r^2+ z_0^2)^{3/2}}\,.
\label{phi_moment}
\end{equation}
Here, ${\bm R}=(x,y,z-z_0)$ is the radius-vector
originating at the source; the first minus sign  is due to the definition
${\bm h}^s=\nabla\,\varphi^s$.  The 2D Fourier transform is:\cite{Grad}
\begin{equation}
\varphi^s({\bm k})  = 2\pi  \,e^{-k z_0}\Big(\mu_z+i\,\frac{
{{\bm\mu}}\cdot{\bm k}}{k}\Big)\,.
\label{FT-mu}
\end{equation}
The response field can now be calculated with the help of Eqs. (\ref{phi_an}).
In general, this can be done numerically; for $\lambda \ll z_0$,
the analytic evaluation is possible.  

 As an example take the moment  directed along $z$ above the flat isotropic
superconducting surface; for the isotropic thin film this problem has been
considered in Ref. \onlinecite{Milosevic}. According to Eq. (\ref{f_sm}) for the
isotropic case
$\varphi^r =
\varphi^s  (1-2 k\lambda)$. Transforming back to real space we obtain:
 \begin{eqnarray}
\varphi^r (r,z) &=& \mu\Big( \frac{Z}{R^3} -2\lambda\,
\frac{2Z^2-r^2}{ R^5}\Big)\,,\nonumber\\
 R &=& \sqrt{r^2+Z^2}\,,\qquad Z=z+z_0\,.
\label{phi(r,z)is2}
\end{eqnarray}
Here, the first term is the field of a moment $- {\bm \mu}$ at
$z=-z_0$, i.e., of the image source. The second term is  the field of the 
proportional to $\lambda$ magnetic quadrupole at the same point. 

 For the magnetic force microscopy, the quantity of interest is the interaction
energy which is given by 
\begin{eqnarray}
{\cal E} &=& -\,{\bm\mu}\cdot {\bm h}^r(0,0,z_0)\label{E}\\ 
&=& -\int \frac{d^2{\bm k}}{(2\pi)^2}\,\varphi^r({\bm
k})e^{-kz_0}(-\mu_zk+i \mu_{\alpha}  k_{\alpha})\nonumber
\end{eqnarray}
Substituting here Eqs. (\ref{f_sm}) and (\ref{FT-mu}) and 
integrating we obtain:
\begin{eqnarray}
{\cal E} &=&\frac{\mu_z^2}{4z_0^3}\Big[1+\frac{3(\lambda_a+\lambda_b)}{4z_0}\Big]
+\frac{\mu_x^2+\mu_y^2}{8z_0^3}\nonumber\\
&+&\frac{3}{32z_0^4}\big[\mu_x^2(3\lambda_b+\lambda_a)+\mu_y^2(3\lambda_a+
\lambda_b)\big]\,.\label{E_tot}
\end{eqnarray} 
In addition to a repulsive force $-\partial{\cal E}/\partial z_0$, the
magnetic moment ${\bm \mu}$ experiences a torque, because the energy depends on
the moment orientation. It is readily seen that if the position and the value of
the magnetic moment are fixed, the minimum of ${\cal E}$ corresponds to  
${\bm \mu}$ situated in the plane $xy$  and parallel to the direction of largest
$\lambda$. The $z$ component of the torque is easily evaluated:
\begin{equation}
\tau_z=\frac{3\mu_{\perp}^2}{16 z_0^4}\,(\lambda_a-\lambda_b)\,\sin 2\beta\,,
\label{torque}
\end{equation}
 where $\mu_{\perp}$ is the in-plane part of the magnetic moment, and $\beta$ is
the angle between ${\bm \mu}_{\perp}$ and $\hat x$. For $\lambda_a=\lambda_b$,
the position of ${\bm \mu}_{\perp}$ in the $xy$ plane is arbitrary, still there
is a torque which tends to rotate ${\bm \mu}$ out of the $z$ direction and to
place it in the $xy$ plane.
 
\subsection{The source as a current loop}

Let the source be a circular current of a radius  $a$ situated in the plane
$z=z_0$. If $a$ and $z_0$ are of the same order of magnitude
(practically, they are both of a few-micron size),  modelling of the
source by a point-size magnetic moment does not suffice. The scalar
potential of the field created by a loop in the plane $z=0$ reads:\cite{LL}   
\begin{equation} 
\varphi^s ({\bm r}) = -\,\frac{I}{c}\int\frac{d{\bm S}\cdot {\bm R}}{R^3}\,.
\label{LL}
\end{equation}
The source current $I$ flows in the counterclockwise direction relative to the
$z$ axis so that an area element $d{\bm S} = dS\,{\hat {\bm z}} $. 
${\bm R}$ is the radius-vector from this element to the interface point $({\bm
r},0)$, and the integral is over the area of the current contour. The position of
the element
$d{\bm S}=d^2{\bm r}^{\prime}{\hat {\bm z}} $ in our situation is $({\bm
r}^{\prime},z_0)$ so that
$ {\bm R} = {\bm r} - {\bm r}^{\prime} - z_0{\hat z}$, and 
\begin{equation} 
\varphi^s ({\bm r},0) = \frac{Iz_0}{c}\int\frac{d^2{\bm r}^{\prime}  
 }{[({\bm r} - {\bm r}^{\prime})^2+z_0^2]^{3/2}}\,.
\label{LL1}
\end{equation}
For the circular loop of the radius $a$, the integral here is over the
circle area. Comparing this with Eq. (\ref{phi_moment}) we see that the
source field can be considered as created by magnetic moments
distributed uniformly over the loop area with the density $I{\hat {\bm z}}/c$ so
that the total moment of the loop is $\pi a^2I{\hat {\bm z}}/c$. 

The 2D Fourier transform of $z_0[({\bm r} - {\bm
r}^{\prime})^2+z_0^2]^{-3/2}$ with respect to ${\bm r}$ is $2\pi
e^{-kz_0}e^{-i{\bm k}\cdot{\bm r}^{\prime}}$, see Eq. (\ref{FT-mu}); therefore,
\begin{eqnarray} 
\varphi^s ({\bm k}) &=& \frac{I}{c}\,2\pi e^{-kz_0}\int d^2{\bm
r}^{\prime} e^{-i{\bm k}\cdot{\bm r}^{\prime}} \nonumber\\ 
&=&  \frac{4\pi^2Ia}{ck} e^{-kz_0}J_1(ka) \,.
\label{ft_loop}
\end{eqnarray}
According to (\ref{BSCCO5}) the 2D Fourier transform of the response potential
for isotropic superconductor is:
\begin{equation} 
\varphi^r ({\bm k}) =   
 \frac{4\pi^2Ia}{ck}\,\frac{q-k}{q+k}\, e^{-kz_0}J_1(ka) \,.
\label{ft_loop_r}
\end{equation}
The $z$-component of the response field follows:
\begin{eqnarray} 
h_z^r ({\bm r},z) =-\, 
 \frac{Ia}{c}\,\int  d^2{\bm k}\frac{q-k}{q+k}\, e^{-k(z+z_0)}J_1(ka)
e^{i{\bm k}\cdot{\bm r}}\nonumber\\
=-\, \frac{2\pi Ia}{c}\,\int_0^{\infty}  dk\,k\,\frac{q-k}{q+k}\,
e^{-k(z+z_0)}J_1(ka)
 J_0(kr)\,.
\label{ft_loop_hz}
\end{eqnarray}

 Further, one can evaluate the response flux through a flat probe  placed above
the superconductor. If the probe is a circular loop of a radius
$a_p$ with the center at ${\bm r}=0$ at the height $z_p\le z_0$, the flux is
given by 
\begin{equation}
\Phi_z^r = -\frac{4\pi^2 Iaa_p}{c}\,\int_0^{\infty}  dk\,
\frac{q-k}{q+k}\, e^{-k(z_0+z_p)}J_1(ka)J_1(ka_p)\,.\label{40}
\end{equation}
All formulas for the isotropic case have been worked out earlier by Clem and
Coffey making use of the cylindrical symmetry of the problem.\cite{Clem}

\subsection{Interaction with vortices}

It is of interest to evaluate the force acting on the vortex tip by the
screening currents in the sample created by a circular current as the field
source; experiments for which this  is relevant are described in Ref.
\onlinecite{Moler-Gardner}. A similar problem for isotropic films and the
magnetic moment as a source has been considered in Ref.
\onlinecite{Milosevic}. We do this for materials isotropic in the $xy$ plane, for
which the force depends only on the distance $r$ from the loop center along with
the loop hight $z_0$. One can calculate
$F_x(x,0;z_0)$ and in the result replace $x$ with $r$:
\begin{eqnarray} 
F_x(x,0) &=&  
 \frac{\phi_0}{c}\,\int_{-\infty}^0  dz j_y = 
\frac{\phi_0}{4\pi}\,\int_{-\infty}^0  dz \,(h_{x,z}-h_{z,x})\nonumber\\
&=&\frac{\phi_0}{4\pi}\,\int \frac{d^2{\bm
k}}{(2\pi)^2}\,e^{ik_xx}\Big(H_x^{(1)}-\frac{ik_x}{q_1}H_z^{(1)}\Big)\,.
\label{Fx1}
\end{eqnarray}
With the help of Eqs. (\ref{BSCCO1}), (\ref{BSCCO2}), and (\ref{ft_loop}) we
  obtain
\begin{equation}
F_r(r) =-\frac{\phi_0 Ia}{c \lambda_{ab}^2}\,\int_0^{\infty} 
\frac{ dk\,k\,e^{-k z_0}}{  q_1(q_1+k) } \,J_1(ka) J_1(kr)\,.
\end{equation}
 
Using $1/\lambda^2=q_1^2-k^2$ and the substitution $t=ka$, we write:
\begin{eqnarray}
F_r  &=&-\frac{\phi_0 I}{2ca}\,\int_0^{\infty} dt\,G\Big({\lambda\over
a}\,t\Big) t\,e^{-z't} J_1(t) J_1( r't) \,,\nonumber\\
G&=& 1-{\lambda\over a}\,t\Big(1+{\lambda^2\over a^2}\,t^2\Big)^{-1/2} ,
\end{eqnarray}
where $z'=z_0/a$ and $r'=r/a$. Usually, the parameter $\lambda/a\ll 1$.
Besides, due to the factor $e^{-z't} J_1(t)$,  the region contributing to the
integral is $0<t<\min (1/z',1)$ because of the oscillating $J_1(t)$ at
large
$t$ (unless $r'\approx 1$). Then, one can expand $G$ in powers of $\lambda t/a$
and keep only the linear term:
\begin{eqnarray}
F_r  &=&-\frac{\phi_0 I}{2ca}\,\Big(f_0-{\lambda\over
a}\,f_1\Big)\,,\nonumber\\ 
f_0&=&\int_0^{\infty} dt\,t\,e^{-z't}J_1(t) J_1( r't) \,,\label{f12}\\ 
f_1&=&\int_0^{\infty} dt\,t^2\,e^{-z't} J_1(t)J_1( r't)\,,\nonumber  
\end{eqnarray}
The integrals here can be expressed in terms of the hypergeometric functions
convenient for the numerical evaluation, see Appendix A.

One can also define a potential energy $U(r)$ such that $F_r=-dU/dr$:
\begin{eqnarray}
U  &=&-\frac{\phi_0 I}{2c}\,\Big(u_0-{\lambda\over
a}\,u_1\Big)\,,\nonumber\\ 
u_0&=&\int_0^{\infty} dt\, e^{-z't}J_1(t) J_0( r't) \,,\label{u12}\\ 
u_1&=&\int_0^{\infty} dt\,t \,e^{-z't} J_1(t)J_0( r't)\,,\nonumber  
\end{eqnarray}

This energy is defined so that $U(\infty)=0$; its value at the origin is
\begin{equation}
U(0) =-\frac{\phi_0
I}{2c}\,\Big[1-\frac{z_0}{(z_0^2+a^2)^{1/2}} -\frac{\lambda
a^2}{(z_0^2+a^2)^{3/2}}\,\Big]\,.\label{U}
\end{equation}
 Thus, the source loop creates a potential well of the depth $|U(0)|$ or a barrier
of the  hight $|U(0)|$ for vortices underneath depending on the current and
vortex directions. This opens an interesting possibility for studying the
behavior of vortices (or antivortices) in tunable potentials.\cite{Moler-Gardner}
In a similar manner, one can evaluate interactions of vortices with other types
of sources.
 
\subsection{Accuracy of the SSM determination of $\lambda$}

Magnetic fluxes of the response field, in particular, $\Phi_z=\int d^2{\bm
r}\,h_z^r({\bm r},z_p)$   (the integral is over the area of a pick-up coil
placed at the hight $z_p$ above the superconducting surface) can be measured with
high accuracy for a given source, given geometry of the coil, and a known height
$z_p$. In principle, this leads to a possibility to measure  the
penetration depth. However, the accuracy of this determination in bound by the
accuracy with which the heights $z_0$ and $z_p$ are known. To demonstrate this
consider the response field of an isotropic material for small $\lambda$'s:
 \begin{eqnarray}
&&h_z^r({\bm r},z_p)  = -\int \frac{d^2{\bm k}}{(2\pi)^2}\,k\varphi^r({\bm
k})e^{i{\bm k}\cdot{\bm r}-kz_p} \,\nonumber\\
&&=-\int \frac{d^2{\bm k}}{(2\pi)^2}\,k\varphi^s({\bm
k} )(1+2\lambda k)e^{i{\bm k}\cdot{\bm r}-kz_p}.    
\end{eqnarray}
If $z_p$ varies by $\delta z_p$, the response field variation is
\begin{equation}
\delta h_z^r =\delta z_p \int \frac{d^2{\bm k}}{(2\pi)^2}\,k^2\varphi^s({\bm
k})(1+2\lambda k)e^{i{\bm k}\cdot{\bm r}-kz_p}\,.
\end{equation}
If only $\lambda$ varies, we have:
\begin{equation}
\delta h_z^r =-2\,\delta\lambda \int \frac{d^2{\bm
k}}{(2\pi)^2}\,k^2\varphi^s({\bm k})(1+2\lambda k)e^{i{\bm k}\cdot{\bm r}-kz_p}\,,
\end{equation}
in other words,
\begin{equation}
\frac{\delta h_z^r}{\delta\lambda} =-2\,\frac{\delta
h_z^r}{\delta z_p} \,.
\end{equation}

Therefore, the accuracy of extracting $\lambda$ from the data on $h_z^r$ cannot
be much better than the knowledge of $z_p$ (the same is true about the source
position $z_0$). The latter is usually known within a fraction of a micron. This
is a severe restriction upon the accuracy of the absolute determination of
$\lambda$.  Still, in principle, SSM allows one to determine accurately the
temperature dependence of $\lambda$ (for  fixed $z_p$ and $z_0$).  

\section{Thin films}

The Meissner response of superconducting thin films can be probed by the SSM
method in yet greater detail than that of bulk samples. The films often
have a large Pearl length $\Lambda=2\lambda^2/d$  ($d$ is the film
thickness) exceeding substantially the size of the SSM sensing loop, making the
SSM measurement to a local probe. Formally, the problem of a thin film in a
field of an external source is simpler than that of the superconducting
half-space, because in the film case there is no ``internal problem" to solve;
instead, the film provides a boundary condition for the outside field
distribution. 

Consider a film in the $x,y$ plane made of a
uniaxial material with the $c$ axis at an angle $\theta$ to the film normal $z$. 
We write the London equation (\ref{general}) for
$i=z$, $h_z-4\pi\lambda^2(m_{xl}j_{l,y}-m_{yl}j_{l,x})/c
=0$, and integrate it over the film thickness:
\begin{equation}
h_z-\frac{2\pi}{c}\Lambda (m_{xx}g_{x,y}-m_{a}g_{y,x}) =0\,,\quad
\Lambda={2\lambda^2\over d},
\label{London_film}
\end{equation}
where $\bm g$ is the sheet current. Note that only two components of the mass
tensor, $m_{xx}=m_a\cos^2\theta+m_c\sin^2\theta$ and
$m_{yy}=m_a$, determine the film anisotropy. 

Using the relation of sheet currents ${\bm g}$ to the tangential components
of the response field,  
\begin{equation}
{2\pi\over c} g_x =  -h_y^r(+0)\,,\,\,\,{2\pi\over c}g_y =h_x^r(+0)  \,, 
\label{e4}
\end{equation}
($+0$ denotes  the upper  film face; the
tangential components satisfy  $h_t^r(+0) =-h_t^r(-0) $), we obtain for $z=+0$: 
\begin{equation} 
  h_z^s+h_z^r +\Lambda (m_{xx}\, h_{y,y}^r  +m_a\, h_{x,x}^r 
 )  =0\,,
\label{L-film}
\end{equation}
   for more detail, see Ref.\onlinecite{KSL}. 
Further, since $h_z^s+h_z^r
=k(\varphi^s-\varphi^r)$ 
we obtain:
\begin{equation}
\varphi^r =\frac{k\varphi^s}{k+ \Lambda
(m_{xx}\,k_y^2 +m_ak_x^2)}  \,,\quad z=+0.
\label{potential-film}
\end{equation}
In particular, for $\theta=0$  we have  
\begin{equation}
\varphi^r
=\frac{\varphi^s}{1+ \Lambda_a
 k}  \,,\qquad \Lambda_a=m_a\Lambda ={2\lambda_{ab}^2\over d}\,.
\label{c||z}
\end{equation}

This result holds also for isotropic materials where $m_a=1$. In this case,
one can readily obtain the field $h_z$ for the circular current  
(\ref{ft_loop}) at the height $z_0$ above the film; its 2D Fourier transform
for $0<z< z_0$ is given by
\begin{eqnarray}
  h_z^r(k,z) &=& -k\varphi^re^{-kz} \nonumber\\&=&- 
\frac{4\pi^2Ia}{c}\, \frac{J_1(ka) }{1+
k\Lambda}\,e^{-k(z+z_0)}\, .
\label{hz}
\end{eqnarray}
This field can be measured by SSM.\cite{John} 

For completeness, we write down also the field on the opposite film side, i.e for
$z<0$, where the response potential is given by 
\begin{equation}
\varphi^r ({\bm r},z<0)=\int\frac{d^2{\bm k}}{(2\pi)^2}\,\varphi^r ({\bm k})\,
e^{i{\bm k}\cdot{\bm r}+kz}\,.
\label{Laplace2}
\end{equation}
The London boundary condition (\ref{L-film}) should now be written in terms of
the field components at $z=-0$:
\begin{equation} 
  h_z^s+h_z^r -\Lambda (m_{xx}\, h_{y,y}^r  +m_a\, h_{x,x}^r 
 )  =0\,,
\label{L-film-}
\end{equation}
which yields the 2D Fourier transform of the response potential
(\ref{potential-film}) with the minus sign. Proceeding as above, we obtain for
the isotropic case:
\begin{equation}
h_z^r(k,z<0) = -
\frac{4\pi^2Ia}{c}\,\frac{J_1(ka) }{1+
k\Lambda}\,e^{k(z-z_0)}\, .
\label{hz}
\end{equation}

The force acting upon a Pearl vortex situated in the film at a radial distance
$r$ from the current ring center is readily evaluated:
\begin{equation} 
F_r(r) =  
 -\frac{\phi_0Ia}{c}\,\int_0^{\infty} \frac{ 
k\,e^{-kz_0}}{1+k\Lambda}J_1(ka)J_1(kr) \,dk\,.
\label{Fr}
\end{equation}

\acknowledgements
These notes were prepared for modeling SSM experiments by John Kirtley and
Kathy Moler; I thank them for encouraging me to put the notes
in a publishable form. Discussions with Walter Hardy, Roman Mints, and Brian
Gardner helped me to focus on calculations which might be useful. In part, the
work was supported by Binational US-Israel Science Foundation. Ames Laboratory is
operated for US DOE by the Iowa State University under Contract No. W-7405-Eng-82.

\appendix

\section{}

The integral $u_0$ of Eq. (\ref{u12}) is known  (see
Ref.\onlinecite{Grad}, 6.612.3):
\begin{eqnarray}
I(r,z) &=& \int_0^{\infty} e^{-zt} J_1(t) J_1( rt)\,dt
\nonumber\\
&=&\frac{1}{2^{5/2}\sqrt{r}u^{3/2}}\,_2F_1\Big({5\over 4},{3\over 4};2;{1\over
u^2}\Big)\,,
\label{I}
\end{eqnarray}
where the primes by $z$ and $r$ are omitted, and  $u=(z^2+r^2+1)/2r $.
Therefore: 
\begin{eqnarray}
&&  \int_0^{\infty}t e^{-zt} J_1(t) J_1( rt)\,dt =-\frac{\partial I}{\partial
z}\label{II}\\ 
&&=\frac{3z r^{-3/2}}{2^{7/2}u^{5/2}}\Big[\, _2F_1\Big({5\over
4},{3\over 4};2;{1\over u^2}\Big)+{5\over 8u^2}\,_2F_1\Big({9\over 4},{7\over
4};3;{1\over u^2}\Big)\Big].\nonumber 
\end{eqnarray}


\begin{references}

\bibitem{Kirtley} J.R. Kirtley,  C.C. Tsuei, M. Rupp,  J.Z. Sun, Lock See
Yu-Jahnes, A. Gupta, M.B. Ketchen, K.A. Moler, and M. Bhushan,
Phys. Rev. Lett. {\bf 76}, 1336 (1996).
 
\bibitem{KirtleyT} J.R. Kirtley, C.C. Tsuei,  K.A. Moler, V.G. Kogan, J.R.
Clem, A.J. Turberfield,
Appl. Phys. Lett. {\bf 74}, 4011 (1999).

\bibitem{Moler-Gardner} B.W. Gardner, J.C. Winn, D.A. Bonn, R. Liang, W.N. Hardy, J.R.
Kirtley, V.G. Kogan, and K.A. Moler,   
Appl. Phys. Lett. {\bf 80}, 1010 (2002).


\bibitem{Clem}J.R. Clem, M.W. Coffey, \prb {\bf 46}, 14662 (1992); M.W. Coffey, 
\prb {\bf 57}, 11648 (1998); M.W. Coffey, \prl {\bf 83}, 1648 (1999).

\bibitem{Milosevic}M.V. Milosevic, S.V. Yampolskii, and F.M. Peters, \prb, {\bf
66}, 174519 (2002).

\bibitem{Obukhov}Yu.V. Obukhov, J. Low Temp. Phys. {\bf 114}, 277 (1999).

\bibitem{K81}  V.G. Kogan, \prb, {\bf 24}, 1572 (1981).

\bibitem{KSL}  V.G. Kogan, A.Yu. Simonov, and M. Ledvij,
Phys. Rev. B {\bf 48}, 392 (1993).

\bibitem{Grad}I.S. Gradshteyn and I.M. Ryzhik, {\it Tables of
Integrals, Series, and Products}, Academic Press, 1980.



\bibitem{John} J.R. Kirtley, unpublished.

\end{references}
\end{document}